\documentclass{emulateapj}
\lefthead{Dewangan et al.}

\slugcomment{Submitted}

\newcommand\asca{{\it ASCA}}

\newcommand\chandra{{\it Chandra}}

\newcommand\xmm{{\it XMM-Newton}}

\newcommand\ks{{\rm~ks}}

\newcommand\kpc{{\rm~kpc}}
\newcommand\mpc{{\rm~Mpc}}

\newcommand\kev{{\rm~keV}}
\newcommand\ev{{\rm~eV}}
\newcommand\kms{\ifmmode {\rm~km\ s}^{-1} \else ~km s$^{-1}$\fi}
\newcommand\Hunit{\ifmmode {\rm~km\ s}^{-1}\ {\rm Mpc}^{-1}
        \else ~km s$^{-1}$ Mpc$^{-1}$\fi}
\newcommand\ctssec{\ifmmode {\rm~count\ s}^{-1} \else ~count s$^{-1}$\fi}
\newcommand\ergsec{\ifmmode {\rm~erg\ s}^{-1} \else
        ~erg s$^{-1}$\fi}
\newcommand\funit{\ifmmode {\rm~erg\ s}^{-1}\;{\rm cm}^{-2} \else
        ~ergs s$^{-1}$ cm$^{-2}$\fi}
\newcommand\phflux{\ifmmode {\rm~photon\ s}^{-1}\;{\rm cm}^{-2}
        \else   ~photon s$^{-1}$ cm$^{-2}$\fi}
\newcommand\efluxA{\ifmmode {\rm~erg\ s}^{-1}\;{\rm cm}^{-2}\;{\rm
        \AA}^{-1} \else ~erg s$^{-1}$ cm$^{-2}$ \AA$^{-1}$\fi}
\newcommand\efluxHz{\ifmmode {\rm~erg\ s}^{-1}\;{\rm cm}^{-2}\;{\rm
        Hz}^{-1} \else ~erg s$^{-1}$ cm$^{-2}$ Hz$^{-1}$\fi}
\newcommand\cc{\ifmmode {\rm~cm}^{-3} \else cm$^{-3}$\fi}
\newcommand\FWHM{\ifmmode {\rm~FWHM} \else ${\rm~FWHM}$\fi}
\newcommand\Msun{\ifmmode M_{\odot} \else $M_{\odot}$\fi}
\newcommand\Lsun{\ifmmode L_{\odot} \else $L_{\odot}$\fi}

\newcommand\hbeta{\ifmmode {\rm H}\beta \else H$\beta$\fi}
\newcommand\Kalpha{\ifmmode {\rm K}\alpha \else K$\alpha$\fi}
\newcommand\nh{\ifmmode N_{\rm H} \else N$_{\rm H}$\fi}
\usepackage{graphicx}
\usepackage{here}

\begin{document}

\title{Evidence for a high energy curvature in the \xmm{} spectrum of
  the ULX NGC~1313~X-1} \author{G. C. Dewangan\altaffilmark{1}, R. E.
  Griffiths\altaffilmark{1}, \& A. R. Rao\altaffilmark{2}}
\altaffiltext{1}{Department of Physics, Carnegie Mellon University,
  5000 Forbes Avenue, Pittsburgh, PA 15213 USA; {\tt email:
    gulabd@cmu.edu, rgriffith@seren.phys.cmu.edu}}
\altaffiltext{2}{Department of Astronomy \& Astrophysics, Tata
  Institute of Fundamental Research, Homi Bhabha Road, Mumbai, 400005
  India; {\tt email: arrao@tifr.res.in}}

\begin{abstract}
  \asca{} X-ray spectra of many ULXs were described in terms of
  optically thick emission from hot ($kT_{in} \sim 1-2\kev$) accretion
  disks, while recent \xmm{} and \chandra{} observations have revealed
  a cool ($kT_{in} \sim 0.2\kev$), soft X-ray excess emission from a
  number of them.  Here we utilize improved calibration and high
  signal-to-noise \xmm{} spectra of NGC~1313 X-1 to present evidence
  for a cool ($\sim 0.2\kev$) soft excess and a curved or a cutoff
  power-law ($\Gamma \sim 1-1.5$, $E_{cutoff} \sim 3-8\kev$).  The
  high energy curvature may also be described by a hot ($\sim
  1-2.5\kev$) multicolor disk blackbody.  The soft excess emission is
  unlikely to arise from a cool disk as its blackbody temperature is
  similar in three \xmm{} observations, despite a change in the
  observed flux by a factor of about two. Thus, previous estimates of
  the black hole mass of $1000\Msun$ for NGC~1313 X-1 based on the
  temperature of the soft excess emission is unlikely to be correct.
  The power-law cutoff energy is found to decrease from $\sim 8\kev$
  to $\sim 3\kev$ when the ULX brightened by a factor of about two.
  The unusual spectral properties of NGC~1313 X-1 are difficult to
  understand in the framework of the disk/corona models generally
  adopted for the black hole binaries or active galactic nuclei and
  may require to invoke super-critical accretion rates.
\end{abstract}


\section{Introduction}
Ultra-luminous X-ray sources (ULXs) are point X-ray sources with
luminosities exceeding $\sim 10^{39}{\rm~erg~s^{-1}}$, outside the
nucleus of a galaxy.  The nature of ULXs continues to be an enigma,
since their adopted isotropic high energy output surpasses the
Eddington limit of even the most massive stellar mass black holes
(BHs), sometimes by large factors.  Several models have been proposed
to explain the high luminosities of ULXs. The most popular is the
``intermediate mass black hole (IMBH)'' with mass $M_{BH} \simeq 10^2
- 10^4\Msun$ (e.g., Colbert \& Mushotzky 1999, hereafter
CM99).  Other models include (i) XRBs with anisotropic emission (King
et al. 2001), (ii) beamed XRBs with relativistic jets directly
pointing towards us i. e., scaled down versions of blazars (Mirabel \&
Rodriguez 1999), and (iii) XRBs with super-Eddington accretion rates
(Begelman 2002).

Several observations suggest that ULXs may be similar or scaled up
versions of the Galactic X-ray binaries (XRBs).  Orbital modulations
in some ULXs (Bauer et al. 2001; Sugiho et al. 2001) imply their
binary nature. Observations of spectral transitions between the low
(hard) and the high (soft) state from two ULXs in IC 342 (Kubota et
al. 2001) demonstrate their similarity to the Galactic BH binaries.
The \asca{} X-ray spectra of ULXs have been described as the emission
from optically thick accretion disks (Makishima et al. 2000 and
references therein). However, the inferred inner disk temperatures in
the range of $\sim 1-2\kev$ are too high for an ULX accreting at
sub-Eddington rates.  Recent \chandra{} and \xmm{} observations of
ULXs have shown soft X-ray excess emission which has been interpreted
as the optically thick emission from thin accretion disks with
temperatures in the range of $\sim 100-300\ev$,suggesting intermediate
mass black holes ($\sim10^3 - 10^4\Msun$) accreting at
sub-Eddington \ ($\sim 0.1L_{Edd}$) rates (see e.g., Miller, Fabian \&
Miller 2004a,b; Miller \& Colbert 2003).  Thus ULXs may either be
$\sim 20-100\Msun$ black holes accreting at near or super Eddington
rates or $\sim 1000-5000\Msun$ black holes with sub Eddington rates.

One of the first ULXs observed by \xmm{} to show the soft excess
emission that led to the $\sim 1000\Msun$ IMBH interpretation, is
NGC~1313 X-1 (Miller et al. 2003).  Here we revisit the spectral
analysis of the ULX NGC~1313 X-1 using all publicly available \xmm{}
observations with improved calibration. We show that the high energy
continuum of the ULX is curved and can be well described either by a
cutoff power law or a high temperature accretion disk component. We
also confirm the soft excess emission earlier detected by Miller et
al. (2003).

\section{Observation and Data Reduction}
The ULX NGC~1313 X-1 has been observed fourteen times by \xmm{} with
exposure times ranging from $9\ks$ to $41\ks$.  We used the EPIC data
obtained from the nine observations that are publicly available and
listed in Table~\ref{tab1}.  The first \xmm{} observation on 2000
October 17 was analyzed by Miller et al. (2003) and led to one of the
first discoveries of a cool soft excess component from an ULX, and a
BH mass estimate of $\sim 10^3\Msun$.  They used the MOS data alone,
as the PN data could not be processed with the SAS 5.3.3 available in
2002. There were also significant differences in the PN and MOS
spectra below $\sim 0.6\kev$: the MOS data showed generally flatter
spectra compared with the PN data.  Here we used SAS 6.5.0 with
updated calibration to process and filter both PN and MOS data
obtained from all the observations listed in Table~\ref{tab1}. All the
observations were affected by a high particle background.  Cleaning of
the flaring particle background resulted in reduced `good' exposure
times listed in Table~\ref{tab1}.  Events in the bad pixels and those
in adjacent pixels were discarded.  Only events with pattern $0-4$
(single and double) for the PN and $0-12$ for the MOS were selected.
Only three observations, with listed exposure times $>6\ks$, are
useful for our purpose of detailed spectral analysis.

\begin{table}
  \centering
  \caption{Log of publicly available \xmm{} observations of NGC~1313. \label{tab1}}
  \begin{tabular}{lcc}
    \tableline\tableline
    Observation  &  Date of &  Usable   \\
    ID       &   Observation  & exposure \\ \tableline
    0106860101 & 2000-10-17 & $23.5\ks$ \\
    0150280101 & 2003-11-25 & $2.5\ks$ \\
    0150280201 & 2003-12-09 & $0.0\ks$ \\
    0150280301 & 2003-12-21 & $7.4\ks$ \\
    0150280401 & 2003-12-23 & $3.2\ks$ \\
    0150280501 & 2003-12-25 & $1.7\ks$  \\
    0150280601 & 2004-01-08 & $6.2\ks$ \\
    0150280701 & 2003-12-27 & $0.0\ks$ \\
    0150281101 & 2004-01-16 & $3.5\ks$ \\ \tableline
  \end{tabular}
\end{table}

\section{Analysis \& Results}
We extracted PN and MOS spectra from all the observations using
circular regions with radii of $35\arcsec$ centered at the position of
NGC~1313 X-1. We also extracted PN and MOS background spectra using
appropriate nearby circular regions free of sources. We created
spectral response files using the SAS tasks {\tt rmfgen} and {\tt
  arfgen}.  Spectral bins were chosen such that there was a minimum of
$20$ and $15$ counts per spectral channel for the PN data of the first
observation and all other data sets, respectively. These spectra were
analyzed with {\tt XSPEC 11.3}. The errors on the best-fit spectral
parameters are quoted at a $90\%$ confidence level.
  
\begin{figure*}
  \centering
  \includegraphics[width=12cm,angle=-90]{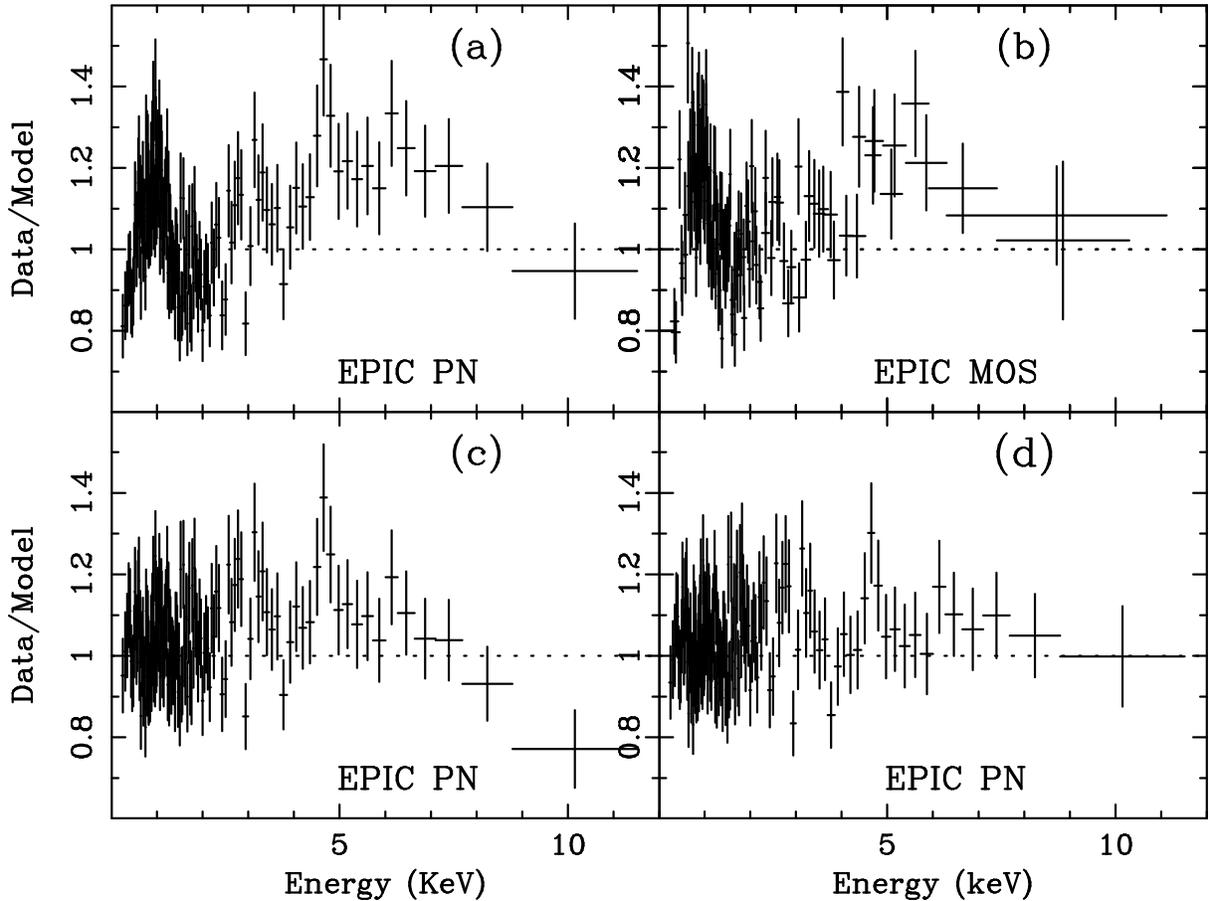}
  \caption{The ratios of observed data obtained on 2000 October and
    the best-fit models. The data are binned heavily for display
    purposes only. Ratios of (a) observed PN data and the best-fit
    absorbed PL model, (b) observed MOS data and the best-fit absorbed
    PL model, (c) the PN data and the best-fit absorbed BB+PL model,
    and (d) the PN data and the best-fit absorbed BB+MCD+PL model. A
    clear cutoff at $\sim 7\kev$ is seen in panel (c).}

  \label{f1}
\end{figure*}

First we fitted a simple absorbed power law (PL) model to the PN and
MOS spectra of NGC~1313 X-1 obtained from the first observation. We
fit the MOS1 and MOS2 data jointly with an overall normalization
constant to account for possible differences in source extraction
areas or calibration uncertainties. We used the $0.2-12\kev$ band in
all the fits.  The simple power law model resulted in minimum $\chi^2
= 749.3$ for $613$ degrees of freedom (dof) and $712.1$ for $656$ dof
for the PN and MOS data respectively, thus providing statistically
unacceptable fits to both the PN and MOS data. We have plotted the
ratios of the PN data and the best-fit PL model in Figure~\ref{f1}a,
and the ratios of the MOS data and the best-fit model in
Fig.~\ref{f1}b. Both the plots are similar and clearly show a broad
hump in the $3-8\kev$ band and a relatively narrow soft hump or excess
emission below $1\kev$. Addition of a soft ($\sim 0.1\kev$) blackbody
(BB) component improves the fit to the PN data significantly ($\Delta
\chi^2 = - 137.8$ for two additional parameters). Similar improvements
are found for the MOS data also. Fig.~\ref{f1}c shows the ratio of the
PN data and the best-fit BB+PL model. The broad hump is now seen as
excess emission in the $\sim 2-7\kev$ band and in the lack of emission
above $\sim 7\kev$. This change in the shape of the broad hump from
Fig.~\ref{f1}a to Fig.~\ref{f1}c is due to the flattening of the
power-law with the addition of the soft blackbody component. Addition
of a multicolor accretion disk blackbody (MCD; {\tt diskbb} in XSPEC)
further improves the fit to the PN data ($\Delta \chi^2 = 16.7$ for
two additional parameters). This is an improvement at a significance
level of $>99.99\%$ based on the maximum likelihood ratio (MLR) test.
To further verify the high energy turnover of the X-ray spectra, we
replaced the PL component in the BB+PL model by a cut-off PL. The BB
{\it plus} cut-off PL model yielded $\Delta \chi^2 = -17.10$ for one
additional parameter as compared to the BB+PL model.  This is an
improvement at a significance level of $>99.99\%$ based on the MLR
test. Thus we conclude that the presence of the broad hump emission is
statistically required.

Since we did not notice significant differences in the PN and MOS
data, we present results based on joint spectral fitting of the PN and
MOS data.  Table~\ref{tab2} lists the best-fit parameters for the
spectral models BB+PL, BB+cutoff PL, BB+MCD and BB+MCD+PL. Good
signal-to-noise spectra are required in order to infer the presence of
the broad hump or high energy turnover. We used only three
observations in which the ULX was detected at a level of $3\sigma$ or
better in the $8-10\kev$ band.  The X-ray spectra of NGC~1313 X-1
obtained on 2000 October 17 and 2003 December 21 clearly statistically
require a cutoff in the power law or a hot MCD component in addition
to the cool soft component, while the lower signal-to-noise spectra of
2004 January 8 are consistent but do not statistically require the
presence of the cutoff or the MCD component.

\begin{figure}
   \centering \includegraphics[width=6cm,angle=-90]{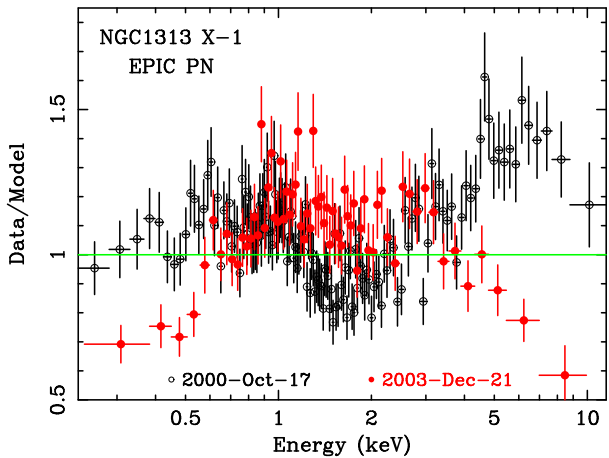}
   \caption{Ratios of the data and the best-fit absorbed power-law
     model jointly fitting the two PN spectra of 2000-10-17 and
     2003-12-21. The absorption columns for the two spectra were tied
     to each other, while the power law indices and normalizations
     were independently. The photon indices are similar ($\Gamma \sim
     2.1$) for both the spectra, despite a factor of two increase in
     the normalization in the observation of 2003 December 21. It is
     clear that the power-law cuts off at lower energy in the 2003
     observation. }
   \label{f2}
 \end{figure}

We show the spectral variability of NGC~1313 X-1 in Figure~\ref{f2} in
terms of the ratios of the data and the best-fit simple absorbed
power-law model derived from a joint fit to the PN data obtained from
the observations of 2000-10-17 and 2003-12-21.  In the joint fit, the
power-law index and the normalization were allowed to vary
independently while the absorption columns for the two spectra were
tied to each other and varied jointly. The best-fit column is $\sim
2.4\times10^{21}{\rm~cm^{-2}}$ and the photon indices are similar
($\Gamma \sim 2.1$) for both the spectra. The ULX was about a factor
of two brighter in the 2003-12-21 observation.  There is a clear
decrease in the power-law cutoff energy (see Fig.~\ref{f2} and
Table~\ref{tab2}).

\section{Discussion}
Based on the spectral analysis of the available high signal-to-noise
PN and MOS data and using the improved calibration of the EPIC
instruments, we found that the X-ray spectrum of NGC~1313 X-1 is
curved at high energies $\ga 6\kev$ and shows a soft excess emission
below $2\kev$.  This ULX showed spectral softening with increasing
flux, similar to that observed from BH XRBs and Seyfert 1 galaxies.
The $0.2-10\kev$ luminosity of the ULX was
$4.3\times10^{39}{\rm~erg~s^{-1}}$ in 2000 October 17 with an increase
to $7.9\times10^{39}{\rm~erg~s^{-1}}$ in 2003 December 21, assuming a
distance of $3.7\mpc$ (Tully 1988).
  
\subsection{A cool corona}
The soft excess and high energy curvature in the X-ray spectra of
NGC~1313 X-1 are well described by a model comprising a cool BB ($kT
\sim 0.2\kev$) and a PL ($\Gamma \sim 1 - 1.5$) with a high energy
cutoff, $E_{cutoff} \sim 3-8\kev$ (see Table~\ref{tab2}). The soft
excess and power-law, with a high energy cutoff, are characteristic
spectral components expected from the accretion disk-corona systems in
BH XRBs and AGNs. The soft excess is thought to be the optically thick
emission from an accretion disk. The cutoff power-law component arises
from thermal Comptonization with a cutoff energy characteristic of the
temperature of the hot comptonizing corona. The observed photon index
of NGC~1313 X-1 is similar to that of BH XRBs in a low/hard state. In
many such BH XRBs and also in Seyfert galaxies, the cutoff energy is
usually observed in the range of $\sim 100\kev$ (see e.g., McClintock
\& Remillard (2004).  If the soft excess and the cutoff power-law
emission from NGC~1313 X-1 arise from a disk-corona, then the corona
must be extremely cool compared to the coronae of both XRB and AGNs.
This is a rather puzzling result. There are also problems with the
interpretation of the soft excess emission as the optically thick
emission from a thin disk (see below).

\subsection{The soft X-ray excess emission}
The temperature of the cool soft component, parameterized as a
blackbody, is $\sim 0.2$ keV in all three observations. The blackbody
normalization increased by a factor of $\sim 2-3$ from 2000 October 17
to 2003 December 21. The cool soft emission was the weakest amongst
the three spectral components in the observation of 2000 October 17,
contributing only $\sim 17\%$ to the $0.2-10\kev$ emission. In the
observations of 2003 December 21 and 2004 January 8, the cool soft
component contributed $\sim 30\%$ and $\sim 6\%$, respectively, in the
$0.2-10\kev$ band. Miller et al. (2003) interpreted this component as
the optically thick, thermal emission from a cool accretion disk and
estimated a black hole mass of $\sim 10^3\Msun$ for NGC~1313 X-1.

Roberts et al. (2005) have pointed out a number of problems with the
above interpretation, which is based on the extrapolation of the X-ray
spectra of XRBs to the IMBH regime and the detection of soft X-ray
excess emission from ULXs. It is also assumed that the accretion disks
of ULXs extend to the last stable orbit, i.e., they are in a 'high'
state. This corresponds to the high/soft state of BH XRBs, where the
optically thick, thermal emission from the disk completely dominates
the $2-10\kev$ spectrum.  This is generally not the case in ULXs,
where the cool soft X-ray excess emission only contributes $\sim
10-30\%$ to the total X-ray emission. The MCD component in BH XRB is
associated with a steep power law (photon index $\Gamma > 2.4$); this
is not always the case in ULXs (see Miller et al. 2003, 2004; Roberts
et al. 2005; Dewangan et al. 2005). The BB+PL model resulted in a
photon index of $1.74\pm0.04$ for NGC~1313 X-1 in the first \xmm{}
observation (see Table~\ref{tab2}). Furthermore, the disk temperature
of BH XRBs increases with their luminosity (see e.g., Miller, Fabian
\& Miller 2004), but the temperature of the soft component from
NGC~1313 X-1 is remarkably constant in spite of the $\sim 80\%$ change
in the $0.2-10\kev$ luminosity. There is also a definite curvature or
cutoff above $\sim 6\kev$ (2000 October) or $\sim 3\kev$ (2003
December) in the X-ray spectrum of NGC~1313 X-1. This curvature is
usually not present in the high state of XRBs: instead, a steep
power-law ($\Gamma \sim 2.5-4$) extending up to several hundred $\kev$
is observed. GRS~1915$+$105 and GRO~J1655-40 are the XRBs that show
e-folding cutoffs at $\sim 3.5\kev$ in their high state (McClintock \&
Remillard 2004), but, unlike NGC~1313 X-1, their power-law slopes are
very steep.

\begin{table}
\centering
\caption{Best-fit spectral model parameters for NGC~1313 X-1.}
\label{tab2}
\begin{tabular}{lccc}
\tableline\tableline
 Model & \multicolumn{3}{c}{Date of Observation} \\
 Parameters  & 2000-10-17 & 2003-12-21\tablenotemark{f} &  2004-01-08 \\ \hline
 BB+PL & \\ \hline
$N_H$($10^{21}{\rm~cm^{-2}}$)&  $2.44_{-06}^{+0.14}$ & $3.7_{-0.4}^{+0.6}$   & $3.0_{-0.2}^{+0.3}$ \\
$kT_{BB}$($\kev$)           & $0.16_{-0.01}^{+0.01}$ & $0.14_{-0.03}^{+0.04}$& $0.18_{-0.04}^{+0.04}$ \\
$A_{BB}$\tablenotemark{a}   & $1.1_{-0.2}^{+0.3}$   &  $2.1_{-1.5}^{+4.5}$  & $1.0_{-0.6}^{+0.4}$ \\
$f_{BB}$\tablenotemark{b}   & $.0.85$             & $1.6$                & $0.85$  \\
$\Gamma$                   & $1.74_{-0.04}^{+0.04}$ & $2.40_{-0.08}^{+0.08}$ & $2.1_{-0.1}^{+0.1}$ \\
$A_{PL}$\tablenotemark{d}  &  $4.8_{-0.3}^{+0.2}$   & $2.1_{-0.2}^{+0.1}$    & $12.4_{-1.6}^{+2.2}$ \\
$f_{PL}$\tablenotemark{b}  &  $3.45$              &  $13.0$               & $7.5$  \\
$f_X$\tablenotemark{e}    &  $2.6$               &  $4.8$               & $4.1$ \\
$L_X$\tablenotemark{e}    &  $4.3$               &  $7.9$                & $6.7$   \\
$\chi^2_{min}$/dof         &   1250.6/1271         & 353.0/372            &  534.6/608  \\ \hline
BB+cutoff PL \\ \hline
$N_H$($10^{21}{\rm~cm^{-2}}$)& $2.0_{-0.3}^{+0.2}$  & $2.8_{-0.4}^{+0.3}$     &  $2.7_{-0.3}^{+0.5}$ \\
$kT_{BB}$($\kev$)          & $0.19_{-0.01}^{+0.02}$ & $0.19_{-0.02}^{+0.04}$  & $0.20_{-0.05}^{+0.04}$ \\
$A_{BB}$\tablenotemark{a}  & $1.0_{-0.1}^{+0.1}$   &  $2.3_{-0.5}^{+0.7}$   &  $1.2_{-0.4}^{+0.5}$ \\
$f_{BB}$\tablenotemark{b}  & $1.0$               & $1.8$                 &  $1.0$ \\
$\Gamma$                  & $1.21_{-0.29}^{+0.12}$ &  $1.38_{-1.13}^{+0.15}$ &  $1.85_{-0.40}^{+0.33}$ \\
$E_{cutoff}$($\kev$)       & $7.8_{-1.8}^{+4.6}$    & $3.6_{-1.7}^{+1.2}$    &  $>0.36$ \\
$A_{PL}$\tablenotemark{d} & $4.0_{-0.5}^{+0.2}$    & $16.9_{-5.0}^{+1.2}$    &  $11.3_{-2.6}^{+2.1}$ \\
$f_{PL}$\tablenotemark{b} &  $2.8$               & $6.8$                  & $6.3$  \\
$f_X$\tablenotemark{e}   & $2.6$                & $4.8$                  & $4.0$ \\
$L_X$\tablenotemark{e}   &  $4.3$               & $7.9$                  & $6.5$ \\
$\chi^2_{min}$/dof        & 1231.7/1270         & $337.2/371$             & $533.6/645$ \\ \hline
BB+MCD \\ \hline
$N_H$($10^{21}{\rm~cm^{-2}}$) & $1.39_{-0.07}^{+0.08}$ & $2.0_{-0.3}^{+0.3}$ & $1.5_{-0.2}^{+0.2}$ \\
$kT_{BB}$($\kev$) &  $0.23_{-0.01}^{+0.01}$ &$0.24_{-0.02}^{+0.02}$ & $0.28_{-}^{+}$ \\
$A_{BB}$\tablenotemark{a} & $1.0_{0.01}^{+0.01}$ &  $2.7_{0.4}^{+0.6}$ & $1.9_{-0.2}^{+0.2}$  \\
$f_{BB}$\tablenotemark{b} & $0.85$ & $2.2$ & $1.6$ \\
$kT_{MCD}$($\kev$) & $2.25_{-0.11}^{+0.09}$ &$1.39_{-0.05}^{+0.14}$ &  $1.88_{-0.17}^{+0.20}$ \\
$A_{MCD}$\tablenotemark{c} & $4.7_{-0.6}^{+1.0}$ & $13.0_{-4.4}^{+4.2}$ \\
$f_{MCD}$\tablenotemark{b} & $2.3$ & $56.2_{-17.8}^{+22.5}$ & \\
$f_X$\tablenotemark{e} & $2.6$  &$4.8$ & $3.3$ \\
$L_X$\tablenotemark{e} &  $4.3$         & $7.9$ & $5.4$ \\
$\chi^2_{min}$/dof &  1259.6/1271 &  344.3/372 &  $547.1/608$ \\ \hline
BB+MCD+PL \\ \hline
$N_H$($10^{21}{\rm~cm^{-2}}$) &  $2.30_{-0.32}^{+0.53}$ &  $2.4_{-0.4}^{+0.5}$ & $2.8_{-1.0}^{+0.9}$ \\
$kT_{BB}$($\kev$) &  $0.18_{-0.02}^{+0.02}$ & $0.21_{-0.03}^{+0.03}$ & $0.19_{-0.05}^{+0.05}$ \\
$A_{BB}$\tablenotemark{a} &   $1.0_{-0.2}^{+0.2}$ & $2.7_{-0.5}^{+0.8}$ & $1.1_{-0.7}^{+2.7}$ \\
$f_{BB}$\tablenotemark{b} & $0.80$ &  $2.2$ & $0.9$  \\
$kT_{MCD}$($\kev$) &  $2.55_{-0.44}^{+0.50}$ & $1.1_{-0.2}^{+0.2}$ & $1.9_{-1.9}^{+6.5}$  \\
$A_{MCD}$\tablenotemark{c} & $1.5_{-0.8}^{+2.7}$ &  $134.1_{-64.6}^{+104.3}$ & $2.0_{-2.0}^{+11.1}$  \\
$f_{MCD}$\tablenotemark{b} &  $1.2$ & $3.8$ & $0.5$  \\
$\Gamma$  &$2.0_{-0.4}^{+0.6}$ &  $1.4_{-3.9}^{+1.0}$ & $2.2_{-1.0}^{+0.8}$ \\
$A_{PL}$\tablenotemark{d} & $3.5_{-1.1}^{+1.1}$ & $1.6_{-1.3}^{+6.8}$ & $11.4_{-9.3}^{+2.5}$ \\
$f_{PL}$\tablenotemark{b} & $2.2$ &$1.4$ & $6.8$  \\
$f_X$\tablenotemark{e} &  $2.6$ & $4.9$ & $4.0$  \\
$L_X$\tablenotemark{e}   & $4.3$ &  $8.0$ & $6.5$ \\
$\chi^2_{min}$/dof & $1231.5/1269$ &  $335.8/370$ &  $534.0/606$ \\ \hline
\end{tabular}
\tablenotetext{a}{BB normalization in units of $10^{-5}\times \frac{L_{39}}{(D10)^2}$, where $L_{39}$ is the source luminosity in units $10^{39}{\rm~erg~s^{-1}}$ and $D10$ is the distance to source in units of $10\kpc$.  }
\tablenotetext{b}{Intrinsic flux in units of $10^{-12}{\rm~ergs~cm^{-2}~s^{-1}}$ and in the $0.2-10\kev$ band.}
\tablenotetext{c}{MCD normalization in units of $10^{-3}(\frac{r_{in}/km}{D10})^2 \times cos{\theta}$, where $r_{in}$ is the inner disk radius and $\theta$ is the disk inclination angle.}
\tablenotetext{d}{Power-law normalization in units of $10^{4}{\rm~photons~keV^{-1}~cm^{-2}~s^{-1}}$ at $1\kev$.}
\tablenotetext{e}{Observed flux and luminosity in units of $10^{-12}{\rm~ergs~cm^{-2}~s^{-1}}$ and $10^{39}{\rm~erg~s^{-1}}$, respectively, and in the band of $0.3-10\kev$.} 
\tablenotetext{f}{Source in MOS chip gaps, only PN data were used.}
\end{table} 

What is the origin of the soft excess component if it is not the
thermal emission from a cool accretion disk?  There are other classes
of accreting objects that show similar soft components.  Many
accreting X-ray pulsars also show soft X-ray excess emission with a
blackbody temperature of $0.1-0.2\kev$ (Paul et al. 2002; Hickox et
al. 2004; Neilsen et al. 2004).  Narrow-line Seyfert 1 galaxies
(NLS1s) and quasars show soft excess emission below $1\kev$ that is
well described by a blackbody. The apparent temperature of the soft
component, $kT \sim 0.1-0.3\kev$, is similar in a diverse selection of
active galactic nuclei (e.g., Czerny et al. 2003; Gierli{\'n}ski \&
Done 2004). This observation can be compared with the similarity of
the temperature of the soft component of NGC~1313 X-1 at different
luminosities. The origin of the soft excess emission is not well
understood. King \& Pounds (2003) suggest that BHs, accreting at or
above the Eddington rate, produce winds that are optically thick in
continuum. Blackbody emission from these winds may provide the
observed soft excess emission from ULXs.

\subsection{Comparison with GRS~1915+105}
\xmm{} observed the Galactic micro-quasar GRS~1915+105 in its
``radio-loud plateau state'' (Martocchia et al. 2005). There is a very
striking similarity between GRS~1915+105 in its plateau state and
NGC~1313 X-1 observed on 2000 October. Both show small amplitude
variability.  Simple absorbed power-law fits to the X-ray spectra of
both the objects result is similar photon index ($\Gamma \sim 1.7$),
the residuals show similar deficits above $8\kev$ and soft excess
emissions around $1\kev$. Fig.~\ref{f1}a is quite similar to Figure~5
in Martocchia et al. (2005).  The above similarity in the X-ray
emission of the XRB and the ULX is observed at a factor of $\sim 10$
difference in luminosities. Thus if the X-ray spectral state is solely
driven by accretion rate relative to the Eddington rate, the BH mass
of NGC~1313 X-1 is likely a factor of $10$ higher ($\sim 140\Msun$)
than that of GRS~1915+105.  The deficit and the soft excess emission
in the spectrum of GRS~1915+105 have been interpreted as the
reprocessing from an ionized disk and reflection from optically thin,
disk wind, respectively.  There are minor differences which, if real,
may lead to the different interpretation of the X-ray spectra and
hence invalidate the direct scaling of their BH masses.  The soft
excess emission of GRS~1915+105 is not well described by a blackbody.
In addition, GRS~1915+105 shows soft X-ray emission lines as well as
iron K$\alpha$ line that are not seen in the spectrum of NGC~1313 X-1.
The ionized disk reflection cannot explain the cutoff or deficit above
$3\kev$ in the 2003-12-21 spectrum.

\subsection{A hot accretion disk}
The X-ray spectra of NGC~1313 X-1 are also well described by the
superposition of a cool BB, a hot MCD and a power-law. However, the
temperature of the MCD component ($kT\sim 2.5\kev$ on 2000 October) is
too high for a thin standard accretion disk.  This is similar to the
``high temperature'' problem of ULXs observed with \asca{} (Mizuno et
al. 1999; Colbert \& Mushotzky 1999; Makishima et al. 2000). One
possible explanation is in terms of a thin disk around a Kerr BH,
since the inner radius of a Kerr disk can be $\equiv 6$ times smaller
than that of a Swarzschild disk, and the temperature of a thin disk
varies as $r_{in}^{-3/4}$, thus resulting in high inner disk
temperatures. Ebisawa et al. (2003) showed that Kerr models work well
for ULXs but the data imply that the disks are highly inclined ($\ga
80^\circ$). This model cannot also explain the soft excess emission
below $2\kev$ from NGC~1313 X-1.

Another attractive scenario is the slim disk model, in which the disk
luminosity can exceed the Eddington luminosity by up to a factor of
$\sim 10$ (Abramowicz et al. 1988, Ebisawa et al. 2003; Kawaguchi
2003). The slim disk is geometrically thick and can be much hotter
than the standard disk.  If this is the case, NGC~1313 X-1 must be
radiating at super-Eddington rates ($\sim 1-10 L_{Edd}$). Its
bolometric luminosity, $L_{bol} \sim 1.5\times
10^{40}{\rm~erg~s^{-1}}$ estimated using the BB+cutoff PL model for
the 2003 December observation, implies a BH mass in the range of $\sim
12 -125\Msun$. A natural consequence of the slim disk model is that
the soft excess emission can be produced in the optically thick winds
resulting from the super-critical accretion rates as discussed above
(Mukai et al. 2003; King \& Pounds 2003).
   
\acknowledgements This work is based on observations obtained with
\xmm{}, an ESA science mission with instruments and contributions
directly funded by ESA Member States and the USA (NASA). This research
has made use of data obtained through the High Energy Astrophysics
Science Archive Research Center Online Service, provided by the
NASA/Goddard Space Flight Center.  GCD acknowledges the support of
NASA award NNG04GN69G.

\end{document}